# Large transport critical currents of powder-in-tube $Sr_{0.6}K_{0.4}Fe_2As_2$ / Ag superconducting wires and tapes

Lei Wang,[1] Yanpeng Qi,[1] Dongliang Wang,[1] Xianping Zhang,[1,2] Zhaoshun Gao,[1] Zhiyu Zhang,[1] Yanwei Ma,[1,*] Satoshi Awaji,[2] Gen Nishijima,[2] and Kazuo Watanabe[2]

[1] Key Laboratory of Applied Superconductivity, Institute of Electrical Engineering, Chinese Academy of Sciences, P. O. Box 2703, Beijing 100190, China

[2] High Field Laboratory for Superconducting Materials, Institute for Materials Research, Tohoku University, Sendai 980-8577, Japan

**Abstract**

We report significant transport critical currents firstly achieved in $Sr_{0.6}K_{0.4}Fe_2As_2$ wires and tapes with a $T_c$ = 34 K, which were fabricated through an in-situ powder-in-tube process. Silver was used as a chemical addition as well as a sheath material. Transport measurements were performed by a standard four-probe resistive method. All the wire and tape samples have shown transport properties. Critical current density $J_c$ was enhanced upon silver addition, and at 4.2 K, a best $J_c$ of ~1200 A/cm$^2$ ($I_c$ = 9 A) was achieved for 20 % silver added tapes, which is the highest in iron-based wires and tapes so far. The $J_c$ is almost field independent between 1 T and 10 T, exhibiting a strong vortex pinning. Such a high transport critical current density is attributed to the absence of reaction layer between the silver sheath and superconducting core, as well as an improved connectivity between grains. We also identify a weak-link behavior from the creep drop of $J_c$ at low fields and a hysteretic phenomenon. Finally, we found that compared to Fe, Ta and Nb tubes, Ag was the best sheath material for the fabrication of high-performance 122 type pnictide wires and tapes so far.

---

[*] Author to whom correspondence should be addressed; E-mail: ywma@mail.iee.ac.cn

**Introduction**

The recent discoveries of superconductivity in LaFeAsO$_{1-x}$F$_x$ (La-1111) and related compounds, with a highest $T_c$ of ~55 K, stimulate worldwide interest in the iron-based system [1-5], however, with much concern about the transport capabilities of such polycrystalline superconductors. [6-11]. The intergranular $J_c$'s of iron-based bulk superconductors have been studied by many groups, and $J_c$'s of 10$^3$-10$^4$ A/cm$^2$ at 4.2 K and self field were reported for Sm- and Nd-1111 bulks, most of which are magnetic results [8-11]. Based on the potential applications, superconductors typically require stabilization using a normal metal cladding for reasons of electrical, thermal, and mechanical protection and, in general, need to be drawn into fine fibers [12]. Indeed, attempts at fabricating iron-based superconducting wires and tapes, through a powder-in-tube (PIT) method, have been made in the 1111,122 and 11 series, however no significant transport critical current has been reported [13-17]. It is presumably due to extrinsic blocking effects, such as non-superconducting phase (wetting phase or disordered phase) at grain boundaries, a dense of array of cracks and a reaction layer between sheath and superconducting core [8-11, 13-15].

The K doped A$_{1-x}$K$_x$Fe$_2$As$_2$ (A=Ba or Sr), exhibiting a high $T_c$ of ~38 K [3], seems more suitable for making superconducting wires or tapes because of the relatively low synthesizing temperature and no oxygen involved, compared with that for the RE-1111 series (RE: rare earth). However, poor transport properties in 122 wires are the principal limitation to technological applications [15]. In order to overcome this problem, we have recently found that the superconducting properties of polycrystalline Sr$_{0.6}$K$_{0.4}$Fe$_2$As$_2$, such as critical transition and irreversibility field, can be improved upon silver addition, probably due to a refined connectivity between grains [18]. In this letter, we introduce silver into Ag-sheathed Sr$_{0.6}$K$_{0.4}$Fe$_2$As$_2$ wires and tapes as a chemical addition. Transport critical currents were measured by a standard four-probe resistive method, and a high transport $J_c$ of ~1200 A/cm$^2$ ($I_c$ = 9 A) at 4.2 K has been achieved in a 20 % silver added tape. These results do support our earlier arguments that the absence of significant transport currents in the previous PIT wires or tapes was caused by extrinsic blocking effects [13-15].

**Experimental details**

$Sr_{0.6}K_{0.4}Fe_2As_2$/Ag/Fe composite wires and tapes were prepared by the in situ powder-in-tube (PIT) process. The details of fabrication process are described elsewhere [19]. Sr filings, Fe powder, As and K pieces, with a ratio Sr : K : Fe : As = 0.6 : 0.4 : 2 : 2.05, were ground in Ar atmosphere for more than 4 hours using ball milling method, with the aim to achieve a uniform distribution. In order to investigate the effect of silver on critical currents, various amounts of metallic silver powder (0-20 wt%) were added in the as-milled mixture. The final powder was filled in a silver tube (OD: 8 mm, ID: 6.4 mm). The composite were filled in an iron tube (OD: 11.6 mm, ID: 8.2 mm). The filled tube was swaged and drawn down to a wire of 1.95 mm in diameter. Some short samples were cut from the as-drawn wires for sintering. The as-drawn wires were subsequently cold rolled into, namely, thick tapes (0.8 mm in thickness) and thin tapes (0.6 mm in thickness). All the wires and tapes were heated at 800-900°C for 35 hours in Ar atmosphere.

Resistivity measurements were carried out by the standard four-probe method using a PPMS system. Dc magnetization measurements were performed with a superconducting quantum interference device SQUID magnetometer. The microstructure was studied using scanning electron microscopy (SEM) after peeling away Ag/Fe sheath. The transport critical currents $I_c$ at 4.2 K and its magnetic dependence were evaluated at the High Field Laboratory for Superconducting Materials (HFLSM) in Sendai, Japan, by a standard four-probe resistive method, with a criterion of 1 $\mu$V cm$^{-1}$. A magnetic field up to 10 T was applied parallel to the tape surface. The $I_c$ measurement was performed for 3-5 samples to check reproducibility.

**Results and discussion**

The choice of a proper sheath was found to be critical in controlling the composition of iron-based superconducting core, and eliminating the reaction layer between superconducting core and sheath. The transverse cross-sections of a typical $Sr_{0.6}K_{0.4}Fe_2As_2$/Ag/Fe wire and tape taken after heat treatments were shown in Fig. 1a. Both Ag/Fe and $Sr_{0.6}K_{0.4}Fe_2As_2$/Ag interfaces were quite clear, indicating silver is benign in proximity to the compound at high temperatures. EDX line-scan has been performed in the direction perpendicular to the longitude of the wires and tapes. It confirms no diffusion of As or Sr into the volumes of Ag and Fe, which benefits superconducting properties of the $Sr_{0.6}K_{0.4}Fe_2As_2$ core. Most importantly, no reaction layer was observed between the silver sheath and the superconducting core (Fig. 1b). By contrast, an early attempt at fabricating $Sr_{0.6}K_{0.4}Fe_2As_2$ wires with a Nb sheath revealed a reaction layer between the sheath and superconducting core after heat treatments, being very harmful to transport current flow [15].

The critical transition temperatures $T_c$ of the samples were determined by the SQUID measurement. The zero-field cooled (ZFC) and field cooled (FC) magnetic susceptibility of the pure and a 20 wt% Ag-added $Sr_{0.6}K_{0.4}Fe_2As_2$ composite tapes were measured under a magnetic field of 10 Oe (Fig. 2). It can be seen that $T_c$ of both samples was estimated to be ~34 K, which is well consistent with that of the $Sr_{0.6}K_{0.4}Fe_2As_2$ bulks and wires prepared previously [15, 18].

$I_c$ measurements of all samples were performed using the standard d.c. four-probe method at 4.2 K in magnetic fields up to 10 T. Zero resistive currents on the current-voltage curves were clearly seen for all wires and tapes. Note that the critical currents of tapes are typically higher than those of corresponding wires, probably due to higher density of superconducting core in tapes, similar to the situation for $MgB_2$ [20]. By contrast, we did not observe significant transport critical currents in Nb or Ta sheathed iron-based wires [13-15], which may be related to bad grain connectivity as well as the reaction layer between the sheath and superconducting core.

Figure 3a shows transport $J_c(H)$ curves for the thick tapes with various amounts of silver addition. Clearly, all silver added samples show a higher $J_c$ than the pure samples in the entire field region, which is in agreement with our earlier observation that the magnetic $J_c$ and irreversibility field $H_{irr}$ can be improved upon silver addition in polycrystalline $Sr_{0.6}K_{0.4}Fe_2As_2$ [18]. The best $J_c$ result was obtained by 20 wt% silver addition, and a $J_c$ of ~500 A/cm$^2$ was reached at 4.2 K and self field, a factor of ~3 higher than that of the pure samples. The almost field independent of $J_c$ between 0.2 and 10 T suggests a strong flux pinning. A steep drop of $J_c$ near 0.2 T was observed for all measured samples, similar to that of sintered YBCO [21], indicating a weak-link behavior.

The field dependence of $J_c$ in an increasing as well as a decreasing field, was also characterized, and a hysteretic phenomenon has been observed. A representative normalized hysteretic $J_c$ curve for a wire sample is shown in the inset of Fig. 3a. After increasing the field monotonically to 8 T, $J_c$ was measured as a function of decreasing field until 0 T. Notable is the increased value of $J_c$ in the region of 0.2-8 T, compared with the data for the virgin curve. The $J_c$ is seen to peak on decreasing the field at 0.4 T, so that for 0 T, the $J_c$ value is substantially reduced as compared to the virgin measurement. The hysteretic effects are supposed to be related to penetration of flux into strong pinning intragranular regions, and that the presence of intragranular critical currents enhances intergranular critical currents when the applied field is reduced from higher values [22]. This phenomenon is also a signature of weak links between grains.

Figure 3b presents transport $J_c$ for pure and 20 wt% Ag-added thin tapes as a function of magnetic field. Two significant observations are recorded. First, a high critical current $I_c$ of 9 A and 0.8 A was achieved for the 20 wt% Ag-added tapes at self field and 10 T, respectively. The original $I$-$V$ plots at 0, 1, 4, 8 and 10 T were shown in the inset of Fig. 3b. Given the average area of superconducting core (~0.75 mm$^2$), a very large critical current density $J_c$ of ~1200 A/cm$^2$ at 0 T and ~100 A/cm$^2$ at high fields was obtained (Fig. 3b). These data is equivalent to the intragrain $J_c$ of RE-1111 estimated from magnetic signals [8-11], and about one order of magnitude higher than the transport $J_c$ for a Fe-Se-Te wire [17]. It should be noted that, a super-current of ~ 1 A still flow in the tape under a high field of 10 T, suggesting a significant vortex pinning supercurrent flowing across grain boundaries. Second, again, the 20 wt% Ag-added thin tapes revealed a higher transport $J_c$ than the pure samples in the entire field region, which is well consistent with the $J_c$(H) curves shown in Fig. 3a.

To clarify the reasons for the enhancement of $J_c$, we studied the differences in the microstructures of the pure and 20 % Ag-added thin tapes. Figure 4 shows typical SEM images of the fractured core layers for the pure and 20 % Ag-added sample. Both samples give a grain size of ~3 μm, while large cracks together with a high porosity were seen for the pure sample (Fig. 4a). In other words, a dense array of extrinsic weak links exists at the grain boundaries, and thus introduces a strong limitation to the flow of currents. In contrast, the 20 % Ag-added samples show a higher density with fewer voids and cracks, resulting in the better connectivity between grains (Fig. 4b). Furthermore, magnified SEM images reveal that, for the pure sample, layered grains stack randomly, thus the grain boundary area accessible for current flow is rather small. But for the 20 % Ag-added sample, most of grains seem to partially melt, resulting in the better connectivity. From the above $J_c$ and structural results, it can be concluded that the silver helps in eliminating cracks and enhancing grain connection, without altering basic grain-boundary properties, because the behavior of magnetic field dependence of $J_c$ is not influenced.

Large transport $J_c$ values obtained in $Sr_{0.6}K_{0.4}Fe_2As_2$ wires and tapes are comparable to those of Sm- and Nb-1111 bulks estimated from magnetic signals [8, 10], in particular one order of magnitude higher than the value obtained in Fe-Se-Te wire [17]. Our work clearly demonstrate the feasibility of fabricating iron-based superconducting wires and tapes with high transport critical currents through the PIT method. Based on the present results, it is believed that further improvement in transport capability of iron-based superconducting wires and tapes is possible upon either achievement of textured materials with clean boundaries [8, 9, 11, 18, 23], or enhancing pinning strengths by the introduction of defect structures [24-25].

**Conclusions**

In summary, we succeeded in eliminating reaction layer by using silver as a sheath material, and achieving significant transport critical currents in $Fe/Ag/Sr_{0.6}K_{0.4}Fe_2As_2$ composite wires and tapes. The transport $J_c$ of $Sr_{0.6}K_{0.4}Fe_2As_2$ can be enhanced upon silver addition, and a transport $J_c$ of ~1200 A/cm$^2$ and ~100 A/cm$^2$ was achieved for a 20 wt% silver added tape at 0 and 10 T, respectively. The $J_c$ enhancement was mainly due to the elimination of cracks and enhanced connectivity. We also identify a weak-link behavior form a creep drop of $I_c$ at low field and a hysteretic phenomenon.

**Acknowledgements**

The authors thank Profs. Haihu Wen, Liye Xiao and Liangzhen Lin for their help and useful discussions. This work was partially supported by the Beijing Municipal Science and Technology Commission under Grant No. Z09010300820907, National Science Foundation of China (grant no. 50802093) and the National '973' Program (grant no. 2006CB601004).

**Captions**

Figure 1 (a) Transverse cross-sections of the Fe/Ag/Sr$_{0.6}$K$_{0.4}$Fe$_2$As$_2$ wire and a typical tape taken after heat treatment. (b) Magnified optical image of the Ag/Sr$_{0.6}$K$_{0.4}$Fe$_2$As$_2$ interface.

Figure 2 Temperature dependence of DC susceptibility of the pure and 20 % Ag-added tapes.

Figure 3 (a) Transport critical current density $J_c$'s of thick tapes (0.8 mm in thickness) with various silver addition from 0 to 20 %. Inset: Hysteresis in a normalized $J_c$. (b) Transport critical current density $J_c$'s of thin tapes (0.6 mm in thickness). Inset: Original I-V plots for the 20 % Ag-added thin tape at 0, 1, 4, 8 and 10 T.

Figure 4 SEM micrographs of superconducting cores of the pure (a, c) and 20 % Ag-added (b, d) thin tapes

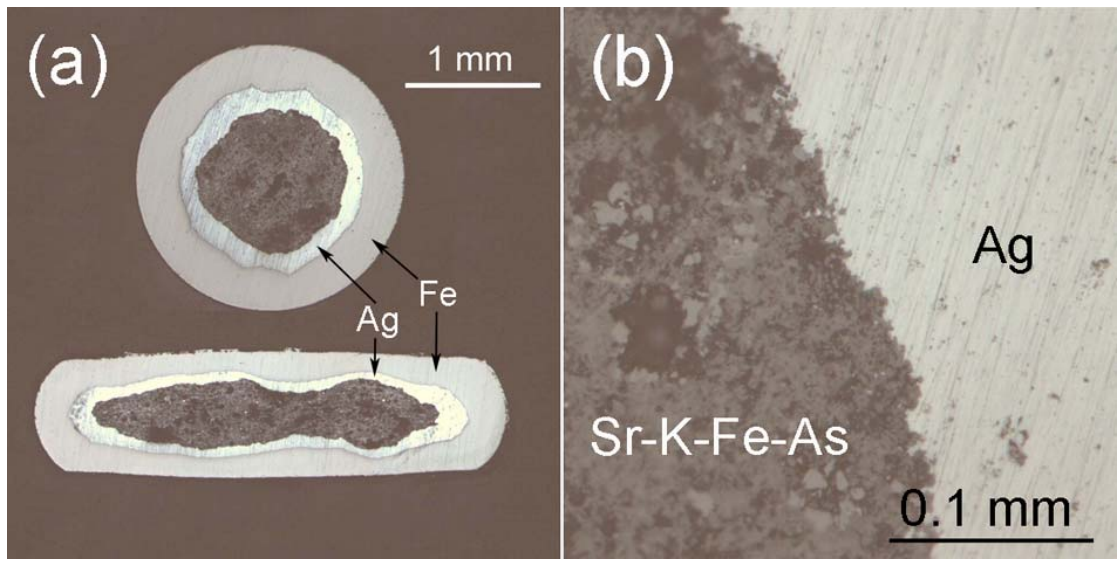

Figure 1 Wang et al.

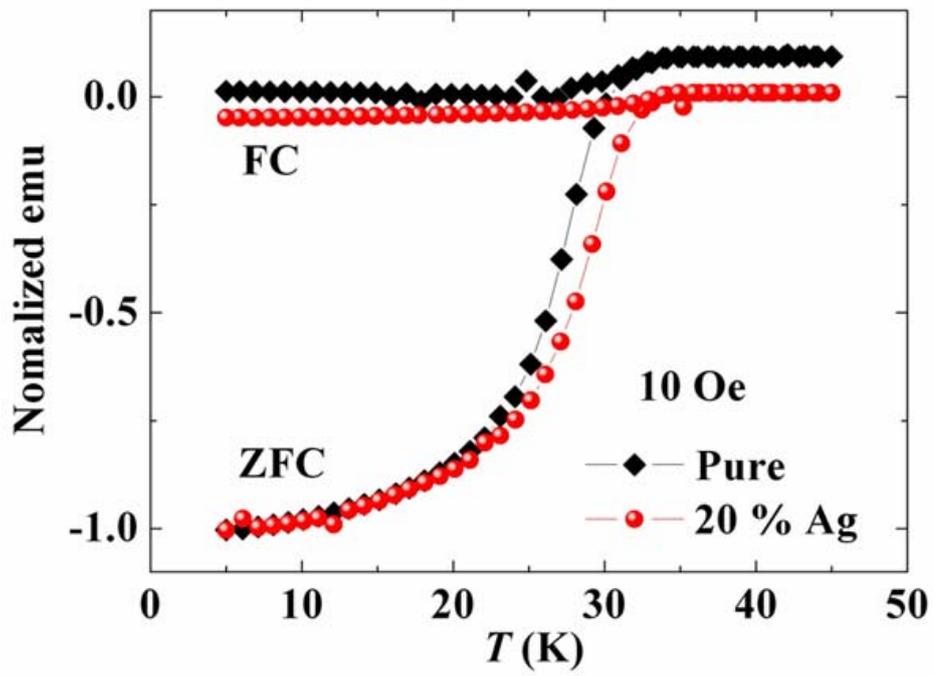

Figure 2 Wang et al.

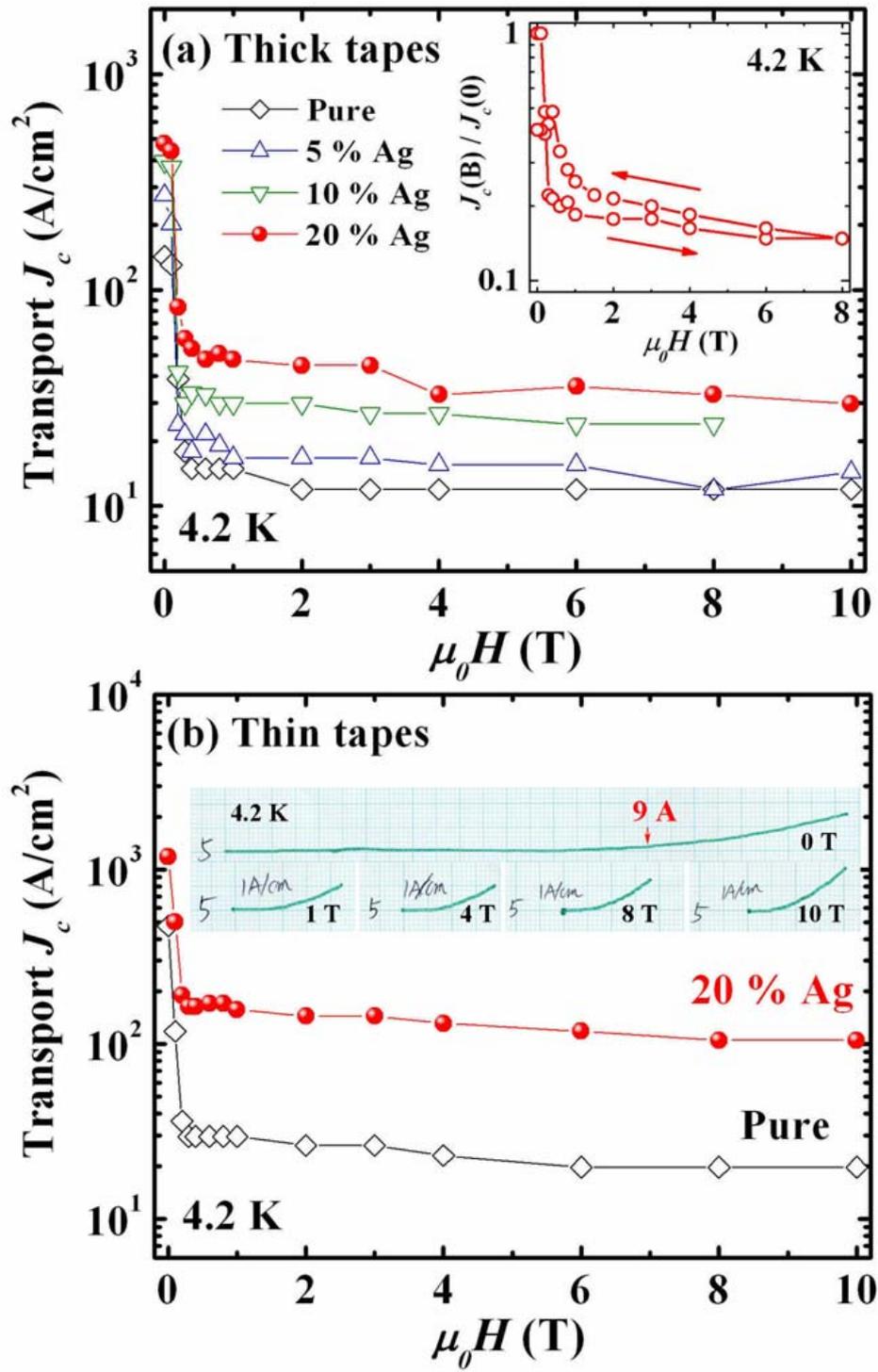

Figure 3 Wang et al.

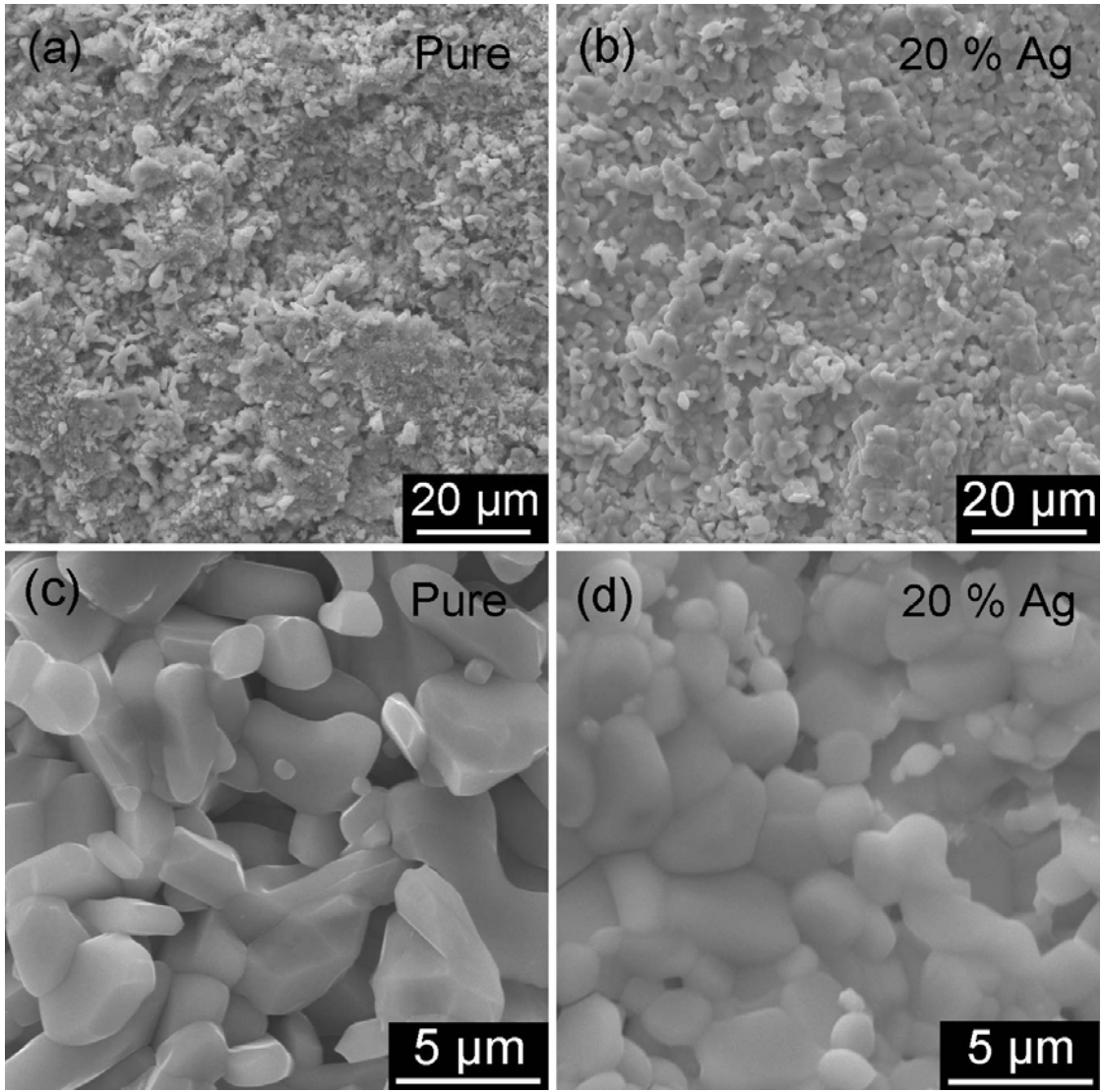

Figure 4 Wang et al.